\documentclass[prl,reprint,superscriptaddress,showpacs]{revtex4-1}
\usepackage[]{graphicx}
\usepackage[]{amsmath}
\usepackage{braket}
\usepackage{color}

\begin{document}

\title{Continuous dynamical decoupling of a single diamond nitrogen-vacancy center spin with a mechanical resonator}

\author{E. R. MacQuarrie}
\affiliation{Cornell University, Ithaca, NY 14853}
\author{T. A. Gosavi}
\affiliation{Cornell University, Ithaca, NY 14853}
\author{S. A. Bhave}
\affiliation{Purdue University, West Lafayette, IN 47907}
\author{G. D. Fuchs}
\email{gdf9@cornell.edu}
\affiliation{Cornell University, Ithaca, NY 14853}

\begin{abstract}
Inhomogeneous dephasing from uncontrolled environmental noise can limit the coherence of a quantum sensor or qubit. For solid state spin qubits such as the nitrogen-vacancy (NV) center in diamond, a dominant source of environmental noise is magnetic field fluctuations due to nearby paramagnetic impurities and instabilities in a magnetic bias field. In this work, we use ac stress generated by a diamond mechanical resonator to engineer a dressed spin basis in which a single NV center qubit is less sensitive to its magnetic environment. For a qubit in the thermally isolated subspace of this protected basis, we prolong the dephasing time $T_2^*$ from $2.7\pm0.1$~$\mu$s to $15\pm1$~$\mu$s by dressing with a $\Omega=581\pm2$~kHz mechanical Rabi field. Furthermore, we develop a model that quantitatively predicts the relationship between $\Omega$ and $T_2^*$ in the dressed basis. Our model suggests that a combination of magnetic field fluctuations and hyperfine coupling to nearby nuclear spins limits the protected coherence time over the range of $\Omega$ accessed here. We show that amplitude noise in $\Omega$ will dominate the dephasing for larger driving fields. 
\end{abstract}

\maketitle

The triplet spin of the nitrogen-vacancy (NV) center in diamond has become a foundational component in both quantum metrology and future quantum information technologies. For sensing, the inhomogeneous dephasing time $T_{2}^*$ of an NV center spin qubit can limit sensitivity to quasi-static fields. For quantum information applications, $T_2^*$ can limit the number and the duration of gate operations that a qubit can undergo. Pulsed dynamical decoupling (PDD) techniques based on the principle of spin echoes refocus inhomogeneous dephasing and can extend $T_2^*$ to the homogeneous spin dephasing time $T_2$ or longer~\cite{childress2006,lange2010,ryan2010,lange2011,naydenov2011,wangPDD2012}. These periodic pulse sequences enable precision sensing and long-lived quantum states, but they come with drawbacks. They usually limit sensing to a narrow bandwidth and erase signal built up from quasi-static fields. Moreover, commuting echo pulses with gate operations makes decoupling during multi-qubit gates a nontrivial task~\cite{vandersar2012}. 

Continuous dynamical decoupling (CDD) offers an alternative method for prolonging $T_2^*$ that can be used when the limitations of PDD become too restrictive. NV center CDD protocols forego the standard Zeeman spin state basis $\{(m_s=)+1,0,-1\}$ in favor of an engineered basis in which the ``dressed'' eigenstates are less sensitive to environmental noise than the bare spin states~\cite{rabl2009,dolde2011,timoney2011,fedder2011,xu2012,hirose2012,mkhitaryan2014,matsuzaki2015,dobrovitski2015}. For an NV center spin qubit, magnetic field fluctuations from nearby paramagnetic impurities and instabilities in a magnetic bias field typically dominate dephasing. A qubit composed of dressed states designed to be more robust to these fluctuations could have a prolonged $T_2^*$ and could be used for precision sensing of quasi-static, non-magnetic fields such as temperature~\cite{awschalomThermo} or strain. For quantum information processing, CDD allows decoupling to continue during gate operations, thus protecting both qubit and gate from dephasing~\cite{timoney2011,xu2012}. 

\begin{figure}[ht]
\begin{center}
\begin{tabular}{c}
\includegraphics[width=\linewidth]{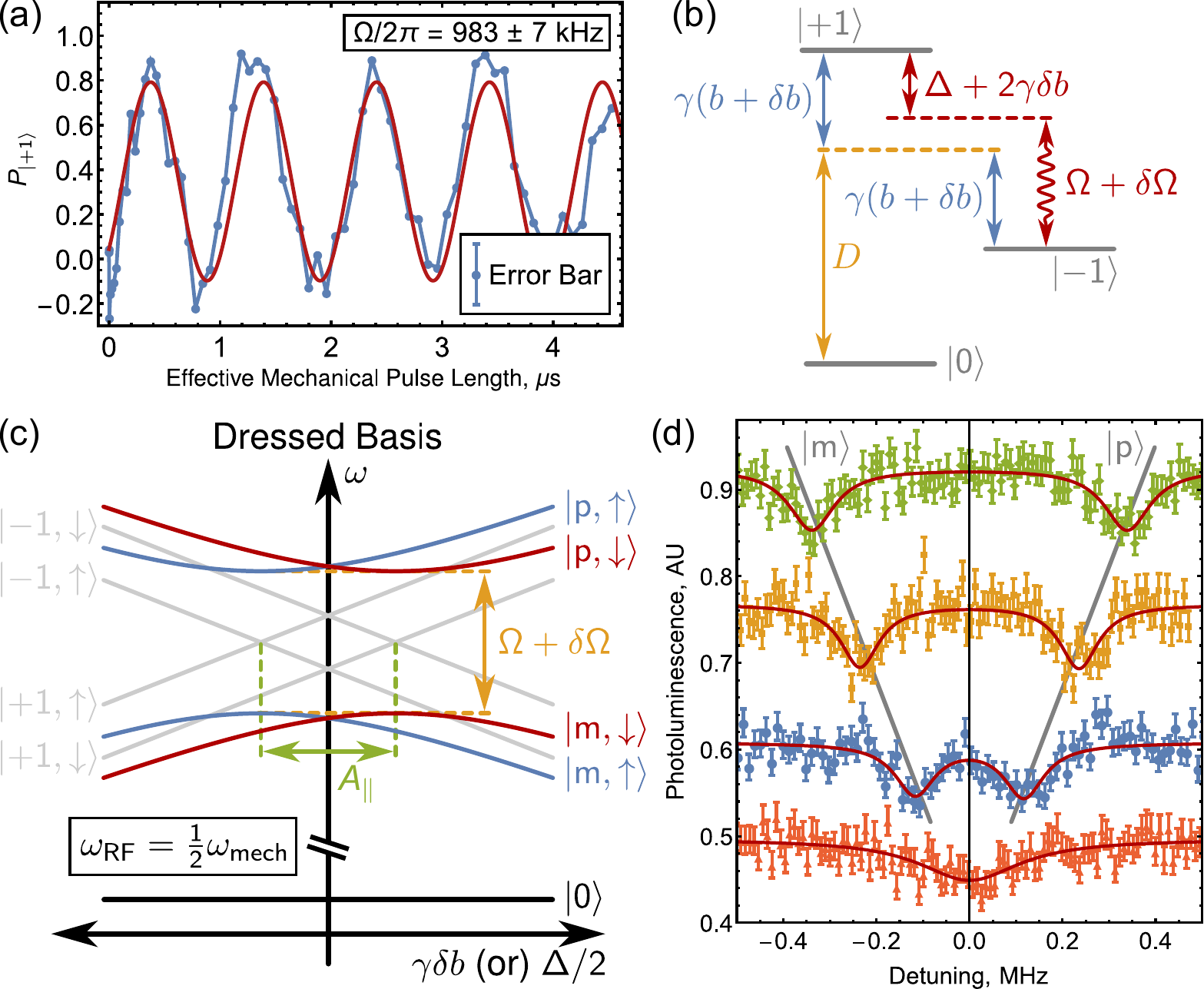} \\
\end{tabular} 
\end{center}
\caption[fig:schematic] {(a) Mechanically driven Rabi oscillations between the $\Ket{-1}$ and $\Ket{+1}$ states of a single NV center within the $m_{I}=+1$ sublevel of the $^{14}$N hyperfine manifold. (b) NV center Zeeman states subject to a static magnetic field $b+\delta b$ and a mechanical driving field $\Omega+\delta\Omega$. (c) Energies of the undressed and dressed eigenstates plotted as a function of both $\gamma\delta b$ and $\Delta/2$ in a reference frame rotating at $\frac{1}{2}\omega_{\text{mech}}$. We include hyperfine sublevels from a nearby $^{13}$C nuclear spin coupled with a strength $A_{\|}$. (d) Spectra of the $\Ket{0}\leftrightarrow\Ket{m}$ and $\Ket{0}\leftrightarrow\Ket{p}$ spin transitions within the dressed state basis. From bottom to top, the mechanical driving fields are $\Omega/2\pi=0$~kHz, $\Omega/2\pi=230\pm10$~kHz, $\Omega/2\pi=470\pm8$~kHz, and $\Omega/2\pi=670\pm10$~kHz.}
\label{fig:schematic}
\end{figure}

Until recently~\cite{barfuss2015}, NV center CDD had only been performed by magnetically driving the $\Ket{0}\leftrightarrow\Ket{+1}$ and $\Ket{0}\leftrightarrow\Ket{-1}$ spin transitions. Advances in diamond mechanical resonator fabrication~\cite{ovartchaiyapongAPL2012,burek2013,MacQuarrie2013,teissier2014,ovartchaiyapong2014} have enabled the use of ac lattice strain to coherently drive the magnetically forbidden $\Ket{+1}\leftrightarrow\Ket{-1}$ spin transition as shown in Fig.~\ref{fig:schematic}a~\cite{MacQuarrie2015,barfuss2015}. Performing mechanical CDD by continuously driving this transition creates a dressed basis that cannot be accessed with conventional magnetic spin control. This basis has eigenstates $\{0,m,p\}$ where $\Ket{m}$ and $\Ket{p}$ are mixtures of only $\Ket{+1}$ and $\Ket{-1}$. The $\Ket{+1}$ and $\Ket{-1}$ states respond diametrically to magnetic fields, making $\Ket{m}$ and $\Ket{p}$ less sensitive to magnetic field fluctuations than their undressed constituents. 

In this work, we perform mechanical CDD to prolong $T_2^*$ of single NV centers and quantify how $T_2^*$ scales with the mechanical dressing field. We determine that, within a thermally isolated subspace of the mechanically dressed basis, a combination of magnetic field fluctuations and coupling to unpolarized nuclear spins limits mechanical CDD over the range of cw dressing fields accessible to our device. Using experiments and theory, we show that for larger driving fields amplitude noise in the mechanical dressing field will become the dominant source of dephasing. 

Compared to magnetic CDD protocols, mechanically dressing the NV center spin has the key benefit that the $\Ket{0}$ state is left unperturbed. This eliminates the need to adiabatically dress and undress the NV center before and after each measurement---a process that can take as long as $50$~$\mu$s each way~\cite{xu2012}. Moreover, the Rabi fields generated by a mechanical resonator are noise filtered above a cutoff frequency $\omega_c$ determined by the quality factor $Q$ and the frequency of the resonance mode $\omega_{\text{mech}}$. This is a valuable feature since driving field noise has previously limited magnetic CDD efforts~\cite{fedder2011,xu2012,cai2012,aiello2013,mishra2014,mkhitaryan2014}. For the resonator used in this work, $\omega_c/2\pi=110$~kHz~\cite{SI}. 

Our derivation of the mechanically dressed energy levels begins in the conventional $\{+1,0,-1\}$ Zeeman basis. As depicted in Fig.~\ref{fig:schematic}b, we consider a static magnetic field $b$ aligned along the NV center symmetry axis that is subject to fluctuations $\delta b$ and a mechanical driving field $\Omega$ that is subject to amplitude fluctuations $\delta\Omega$. We work within the $m_I=+1$ sublevel of the $^{14}$N hyperfine manifold. 
In diamonds with a natural distribution of carbon isotopes, nearby $^{13}C$ nuclear spins typically couple to the NV center spin. Weak coupling to a single $^{13}$C spin is described by the hyperfine perturbation $H_{C}=A_{\|}S_{z}I_{z}$ where $S_z$ and $I_z$ are the spin-$1$ and spin-$\frac{1}{2}$ Pauli matrices, respectively, and $A_{\|}$ is the coupling strength~\cite{slichter}. Applying the rotating wave approximation, we transform into the reference frame rotating at $\frac{1}{2}\omega_{\text{mech}}=\frac{1}{2}(2\gamma b+\Delta)$ where $\Delta$ gives the detuning of $\omega_{\text{mech}}$ from the $\Ket{+1}\leftrightarrow\Ket{-1}$ spin state splitting. Diagonalizing the resulting Hamiltonian gives eigenstates $\{0,m,p\}$ with energies $\{-D,-\frac{1}{2}\sqrt{(\Omega+\delta\Omega)^2+\xi_{\pm}^2}, \frac{1}{2}\sqrt{(\Omega+\delta\Omega)^2+\xi_{\pm}^2}\}$ where $\xi_{\pm}\equiv \Delta+2\gamma\delta b\pm A_{\|}$ for the $m_{I}=\pm\frac{1}{2}$ sublevel of the $^{13}$C manifold. Here, $\gamma/2\pi=2.8$~MHz/G is the NV center gyromagnetic ratio and $D\simeq D_0+\frac{dD}{dT}\Delta T$ is the zero-field splitting where $D_0/2\pi=2.87$~GHz and $\frac{dD}{dT}=-74\times 2\pi$~kHz/$^{\circ}$C is the temperature dependence of $D$~\cite{doherty2012,awschalomThermo,budkerThermo,SI}. 

Fig.~\ref{fig:schematic}c plots the energy levels of the dressed and undressed eigenstates as a function of both $\gamma\delta b$ and $\Delta/2$. The Larmor frequency $\omega_{i,j}$ at which a qubit accumulates phase is given by the energy splitting between the $\Ket{i}$ and $\Ket{j}$ qubit states. Variations in $\delta b$ will cause $\omega_{i,j}$ to fluctuate in time, dephasing the qubit. Mechanically dressing the NV center opens an avoided crossing between the $\Ket{m}$ and $\Ket{p}$ states at $\gamma\delta b=\frac{1}{2}(\Delta\pm A_{\|})$, which reduces the sensitivity of $\omega_{i,j}$ to variations in $\delta b$ and protects the qubit from dephasing. 

We spectroscopically observe this manufactured avoided crossing by first tuning the $m_{I}=+1$ $^{14}N$ sublevel of the $\Ket{+1}\leftrightarrow\Ket{-1}$ splitting into resonance with the $\omega_{\text{mech}}/2\pi=586$~MHz mechanical mode of a high-overtone bulk acoustic resonator (HBAR)~\cite{MacQuarrie2013,SI}. With $\Omega$ resonantly addressing this transition, we sweep the detuning of a $\Omega_{\text{mag}}/2\pi\sim80$~kHz magnetic driving field through the resonance of the undressed $\Ket{0}\leftrightarrow\Ket{-1}$ transition. The resulting spectra are shown in Fig.~\ref{fig:schematic}d for several values of $\Omega$. We interleave measurements of the dressed and undressed spectra to simultaneously measure $\omega_{0,-1}$; $\omega_{0,m}$; and $\omega_{0,p}$. The relation $\frac{1}{2}(\omega_{0,m}+\omega_{0,p})-\omega_{0,-1}=\frac{1}{2}\Delta$ then provides a means to more precisely zero $\Delta$~\cite{SI}. By operating at $\Delta=0$ where $\frac{\partial\omega_{i,j}}{\partial\delta b}\Bigr|_{\Delta=0}\neq 0$, we detune $\Omega$ equally from each $^{13}$C sublevel. This dresses both sublevels equivalently, preserving the full spin contrast of our measurements and maintaining the $^{13}$C manifold as a degree of freedom. Alternatively, we could maximally protect one nuclear sublevel at the expense of the other by operating at $\Delta=\pm A_{\|}$ where $\frac{\partial\omega_{i,j}}{\partial\delta b}\Bigr|_{\Delta=\pm A_{\|}}=0$ for one of the two sublevels. For an unpolarized $^{13}$C spin, however, such a strategy would halve the measured spin contrast, limiting the utility of mechanical CDD. 

Next, we perform Ramsey measurements within the dressed basis to quantify the decoherence protection offered by mechanical CDD. We begin by examining the qubit derived from the $\{0,p\}$ subspace, which is minimally perturbed from the more familiar $\{0,-1\}$ qubit. For these measurements, a $\Omega_{\text{mag}}/2\pi=696\pm 7$~kHz magnetic $\pi/2$-pulse resonant with the $\Ket{0}\leftrightarrow\Ket{p}$ transition populates the $\{0,p\}$ subspace. Because $\Omega_{\text{mag}}>\omega_{m,p}$, the $\{0,m\}$ subspace is also populated. A second magnetic $\pi/2$-pulse of the same strength returns the spin population to $\Ket{0}$ for optical readout. We advance the phase of the second pulse by $\omega_{\text{rot}}\tau$ to help visualize the decay. We then repeat this protocol as a function of $\Omega$~\cite{SI}. 

\begin{figure}[ht]
\begin{center}
\begin{tabular}{c}
\includegraphics[width=\linewidth]{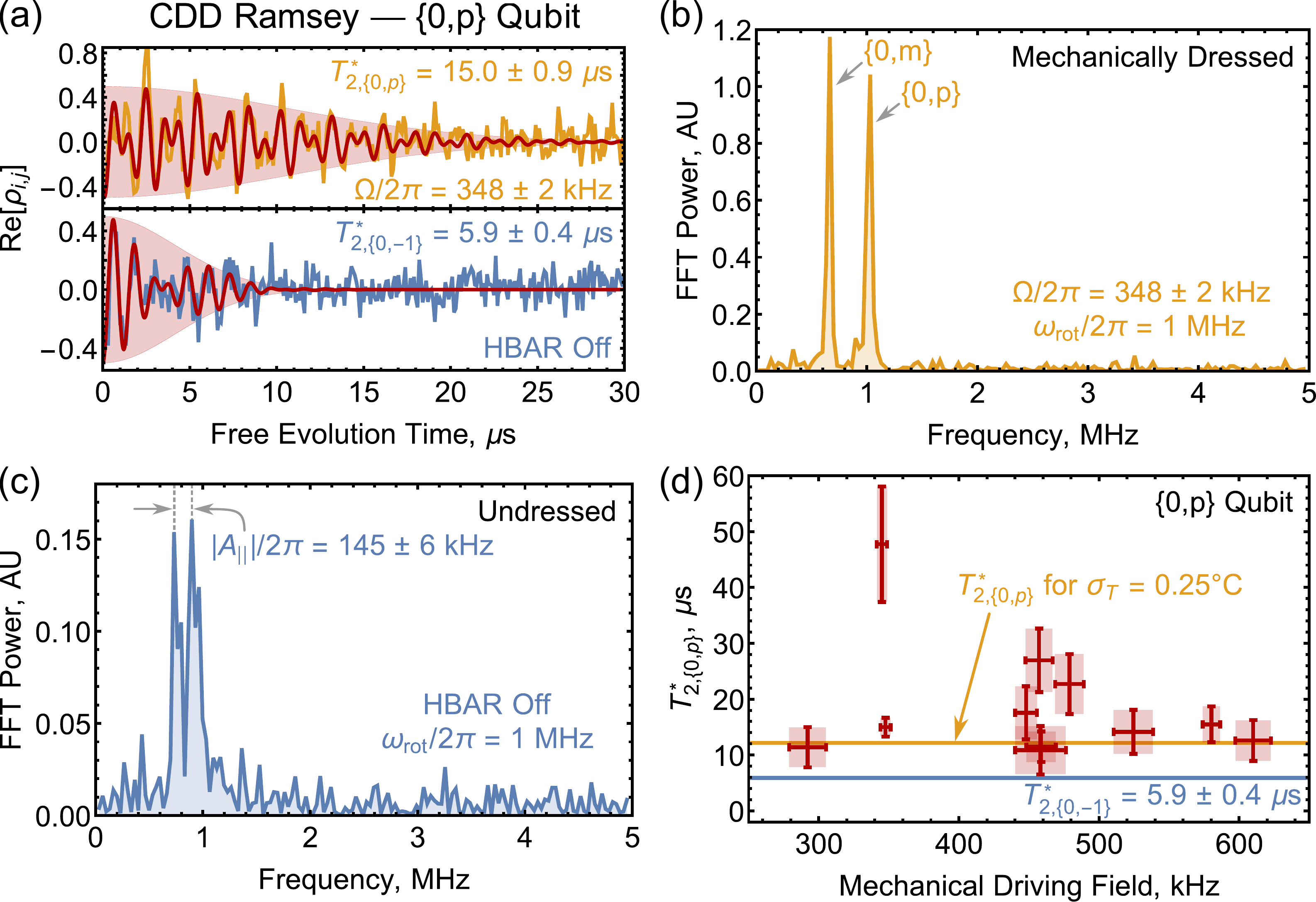} \\
\end{tabular} 
\end{center}
\caption[fig:fig0p] {(a) Ramsey measurements of the $\{0,-1\}$ qubit and the $\{0,p\}$ qubit protected by a $\Omega/2\pi=348\pm2$~kHz mechanical dressing field. (b,c) Fourier spectra of the Ramsey data in (a). (d) Coherence time of the $\{0,p\}$ qubit plotted as a function of $\Omega$. Error bars in (d) indicate $95\%$ confidence intervals. }
\label{fig:fig0p}
\end{figure}

Fig.~\ref{fig:fig0p}a shows that a $\Omega/2\pi=348\pm4$~kHz dressing field extends $T_2^*$ from $5.9\pm0.4$~$\mu$s to $15.0\pm0.9$~$\mu$s. We approximate the decay of our CDD Ramsey signal with a Gaussian envelope. This is not strictly correct because $\omega_{i,j}$ varies non-linearly with fluctuations in the environment. Nevertheless, when $\frac{\partial\omega_{i,j}}{\partial\delta b}\neq 0$ Gaussian decay reasonably approximates the dephasing over the range of $\Omega$ employed in this work and facilitates comparison with the undressed qubit coherence~\cite{SI}. Fig.~\ref{fig:fig0p}b,c provide the Fourier spectrum of each measurement in Fig.~\ref{fig:fig0p}a. Beating in the undressed Ramsey signal reveals a $|A_{\|}|/2\pi=145\pm 6$~kHz coupling to a nearby $^{13}$C spin.

If the $\{0,p\}$ qubit coherence is limited by $\delta b$, then $T_{2,\{0,p\}}^*$ should scale linearly with $\Omega$. However, as Fig.~\ref{fig:fig0p}d shows, plotting $T_{2,\{0,p\}}^*$ as a function of $\Omega$ reveals an erratic distribution with a clustering around $T_{2,\{0,p\}}^*\sim 12$~$\mu$s. By monitoring the temperature of our sample over the course of several measurements, we identified that this effect arises from long-term temperature instabilities~\cite{SI}. Temperature enters the dressed NV center Hamiltonian through the zero-field splitting $D$, which varies at a rate of $\frac{dD}{dT}=-74\times 2\pi$~kHz/$^{\circ}$C~\cite{awschalomThermo,budkerThermo} and contributes to $\omega_{0,p}$ and $\omega_{0,m}$. Gaussian thermal drift with a standard deviation of $\sigma_{T}=0.25^{\circ}$C will dephase the $\{0,p\}$ qubit in $T_{2,\{0,p\}}^*=\frac{\sqrt{2}}{\sigma_{T} dD/dT}=12$~$\mu$s. Coherence times measured during periods of minimal thermal drift exceed this limit, indicating that mechanical CDD isolates the $\{0,p\}$ qubit from magnetic noise more successfully than Fig.~\ref{fig:fig0p}d implies. Thermal instabilities take over as the dominant dephasing channel, however, which suggests mechanical CDD could offer an alternative thermometry protocol to thermal CPMG~\cite{awschalomThermo}.

With the $\{0,p\}$ qubit subdued by thermal fluctuations, we turn to the $\{m,p\}$ qubit to fully explore the efficacy of mechanical CDD at enhancing $T_2^*$. The Larmor frequency $\omega_{m,p}$ is independent of $D$, making the $\{m,p\}$ qubit insensitive to changes in temperature. To measure $T_{2,\{m,p\}}^*$, we populate the $\{m,p\}$ subspace with a magnetic double quantum (DQ) $\pi$-pulse of frequency $\omega_{DQ}=\frac{1}{2}(\omega_{0,m}+\omega_{0,p})$ and strength $\Omega_{\text{mag}}/2\pi=1513\pm8$~kHz~\cite{mamin2014}. After a free evolution time $\tau$, a second DQ $\pi$-pulse of the same strength transfers the spin back to the $\Ket{0}$ state, where fluorescence readout measures the $\{m,p\}$ qubit coherence~\cite{SI}. For these measurements, we studied a second NV center located nearby the NV center that was used in the $\{0,p\}$ qubit measurements. Both NV centers are quantitatively similar and have comparable $T_{2,\{0,-1\}}^{*}$ and $A_{\|}$. 

\begin{figure}[ht]
\begin{center}
\begin{tabular}{c}
\includegraphics[width=\linewidth]{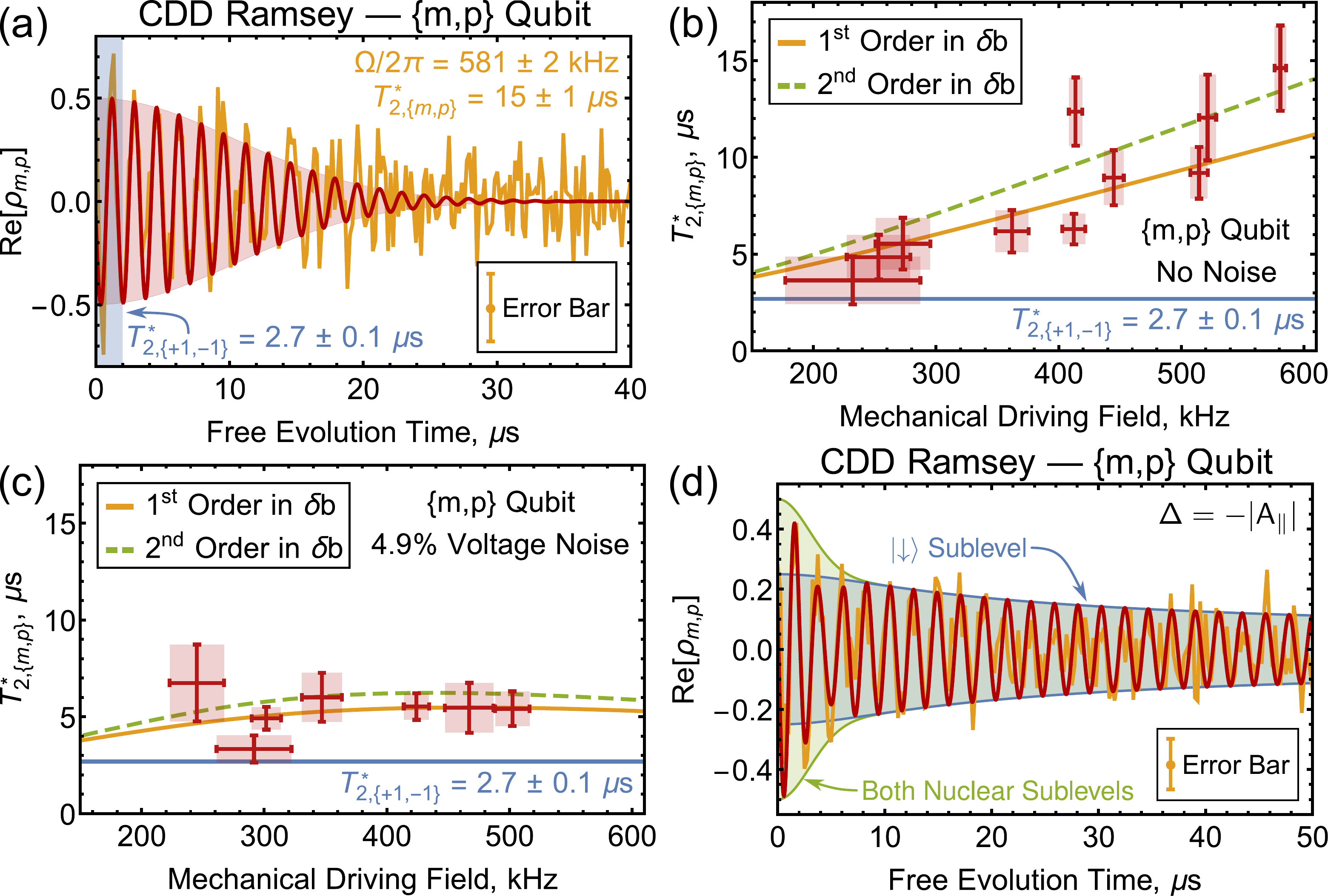} \\
\end{tabular} 
\end{center}
\caption[fig:figMP] {(a) Ramsey measurements of the $\{m,p\}$ qubit protected by a $\Omega/2\pi=581\pm2$~kHz mechanical dressing field. (b,c) Coherence time of the $\{m,p\}$ qubit plotted as a function of $\Omega$ for measurements where $\Omega$ was (b) actively stabilized and (c) given a Gaussian noise profile. Error bars in (b,c) indicate $95\%$ confidence intervals. (d) Ramsey measurement of the $\{\Ket{m,\downarrow},\Ket{p,\downarrow}\}$ qubit protected by a $\Omega/2\pi=455.7\pm0.5$~kHz mechanical dressing field under the condition $\Delta=-|A_{\|}|$. }
\label{fig:figMP}
\end{figure}

Fig.~\ref{fig:figMP}a shows a typical CDD Ramsey measurement for the $\{m,p\}$ qubit. The undressed analog of the $\{m,p\}$ qubit is the $\{+1,-1\}$ qubit, and its $T_{2,\{+1,-1\}}^*=2.7\pm0.1$~$\mu$s coherence time is indicated by the shaded region in Fig.~\ref{fig:figMP}a. A $\Omega/2\pi=581\pm2$~kHz dressing field extends the $\{m,p\}$ qubit coherence to $T_{2,\{m,p\}}^*=15\pm1$~$\mu$s~\cite{SI}. 

In order to quantitatively study how the measured spin protection scales with $\Omega$, we examine quasi-static deviations in $\omega_{i,j}$~\cite{ithier2005}. Because we work in a reference frame rotating at $\frac{1}{2}\omega_{\text{mech}}$, low frequency electric and strain field noise are averaged away, and---as noted above---the $\{m,p\}$ qubit is isolated from thermal noise. We thus examine dephasing from only two independent sources: $\delta b$ and $\delta\Omega$. 

Consider a generic deviation of the form $\delta\omega_{i,j}=\alpha\delta x$ where $\alpha$ is a constant and the fluctuation $\delta x$ follows a Gaussian distribution with standard deviation $\sigma_x$. The associated dephasing rate is $\Gamma_x=\sqrt{2}\pi\alpha \sigma_x$, and the dephasing time from a collection of uncorrelated noise sources is given by $\frac{1}{T_2^*}=\frac{1}{2\pi} \sum\Gamma_{i}$. Assuming that $\delta b$ dominates dephasing of the $\{0,-1\}$ qubit, we then find $\gamma\sigma_b/2\pi=(\sqrt{2}\pi T_{2,\{0,-1\}}^*)^{-1}=42\pm2$~kHz where $T_{2,\{0,-1\}}^*=5.4\pm0.3$~$\mu$s for this NV center. For the $\{m,p\}$ qubit, expanding $\omega_{m,p}$ to first order in $\delta b$ gives the dephasing rate from magnetic fluctuations to leading order in $\delta b$ as $\Gamma_b=2\sqrt{2}\kappa|A_{\|}|/T_{2,\{0,-1\}}^*$ where $\frac{1}{\kappa}\equiv\frac{1}{\sqrt{2}\pi}\sqrt{A_{\|}^2+\Omega^2}$. Similarly, expanding $\omega_{m,p}$ to first order in $\delta \Omega$ gives the dephasing rate due to fluctuations in the amplitude of $\Omega$ as $\Gamma_\Omega=\kappa\Omega\sigma_{\Omega}$~\cite{SI}. 

Our measurements of $T_{2,\{m,p\}}^*$ employ a feedback protocol to level the power supplied to the HBAR and reduce $\delta\Omega$ to $\sim0.03\%$ of $\Omega$. For the range of $\Omega$ accessed here, this level of stability makes $\Gamma_{\Omega}\ll\Gamma_b$, and we can ignore the effects of $\delta\Omega$. To first order in $\delta b$, the dephasing time of the $\{m,p\}$ qubit is then given by $T_{2,\{m,p\}}^*=\frac{2\pi}{\Gamma_b}$.

Fig.~\ref{fig:figMP}b plots $T_{2,\{m,p\}}^*$ as a function of $\Omega$. We attribute scatter in the data mainly to deviations from the $\Delta=0$ condition. For $\Omega\lesssim 10\gamma\sigma_b=(420\pm 20)\times 2\pi$~kHz, the first order expansion in $\delta b$ correctly predicts $T_{2,\{m,p\}}^*$. However, as $\Omega$ increases and $\frac{\partial\omega_{m,p}}{\partial\delta b}\Bigr|_{\Delta=0}$ diminishes, the measured coherence times begin to surpass the predictions of the first order model. To account for this, we extend our model to second order in $\delta b$ and numerically solve the resulting non-Gaussian decoherence envelope for the $\frac{1}{e}$ decay time~\cite{SI,ithier2005}. As seen in Fig.~\ref{fig:figMP}b, the model that corrects to second order in $\delta b$ more accurately predicts $T_{2,\{m,p\}}^*$ for $\Omega\gtrsim 10\gamma\sigma_b$. This suggests that for these higher dressing fields, the $\{m,p\}$ qubit coherence remains limited by $\delta b$. The cw power handling capabilities of our device prohibited measurements at larger $\Omega$, but these results indicate that $T_{2,\{m,p\}}^*$ would continue to increase with $\Omega$. 

To test the predictive capabilities of our model, we intentionally increase $\delta\Omega$ to the point where $\Gamma_{\Omega}$ becomes the dominant dephasing channel. To do this quantitatively, we monitor the voltage reflected from the HBAR $V_R$, which scales linearly with $\Omega$. We then periodically randomize the power supplied to the HBAR to give $V_R$ a Gaussian distribution with standard deviation $\sigma_V=\eta \langle V_{R}\rangle$ where $\eta$ is a constant. This yields a Gaussian distribution of $\Omega$ with a standard deviation $\sigma_{\Omega}=(\langle\Omega\rangle+\alpha)\eta$ where $\alpha/2\pi=-133\pm7$~kHz is a constant related to our measurement of $V_R$~\cite{SI}. The dephasing time is then given by $T_{2,\{m,p\}}^*=\frac{2\pi}{\Gamma_b+\Gamma_\Omega}$. 

Fig.~\ref{fig:figMP}c shows the measured and predicted $T_{2,\{m,p\}}^*$ for $\eta=4.9\pm0.2\%$. The decoherence in these measurements is dominated by $\delta\Omega$. Therefore, the model accurately predicts $T_{2,\{m,p\}}^*$ whether $\Gamma_b$ is correct to first or second order in $\delta b$. Power leveling can effectively zero $\delta\Omega$ over the range of $\Omega$ measured here, but these results suggest that in a more efficient device where a larger $\Omega$ is attainable, amplitude noise would eventually limit the protection that mechanical CDD offers in the power-leveled case. 

We conclude by maximally protecting the $\Ket{\downarrow}$ $^{13}$C sublevel of the $\{m,p\}$ qubit at the expense of the $\Ket{\uparrow}$ sublevel to examine the limits of mechanical CDD. By setting $\Delta=-|A_{\|}|$ where $|A_{\|}|/2\pi=150\pm4$~kHz for this NV center, we establish the condition $\frac{\partial\omega_{m,p}}{\partial\delta b}\Bigr|_{\Delta=-|A_{\|}|}=0$ for the $\Ket{\downarrow}$ sublevel. To second order in $\delta b$, the coherence of this sublevel is then described by
\begin{equation}
\text{Re}[\rho_{m,p}]=\frac{a}{4}\sqrt{\frac{\Omega}{\sqrt{\Omega^2+(2\gamma\sigma_b)^4\tau^2}}}\cos\left[\Omega\tau+\phi\right] +c
\label{eq:highCohere}
\end{equation}
where $a$ accounts for imperfect spin contrast, $c$ is a constant background, and $\phi$ is a constant phase. The result of this measurement for a $\Omega/2\pi=455.7\pm0.5$~kHz dressing field is shown in Fig.~\ref{fig:figMP}d. The data have been fit to a sum of Eq.~\ref{eq:highCohere} and Gaussian decay of the $\Ket{\uparrow}$ coherence where only $\Omega$, $\phi$, $c$, and $T_{2,\uparrow}^*$ were allowed to vary as free parameters~\cite{SI,ithier2005}. As the shaded regions of the figure highlight, the $\Ket{\uparrow}$ sublevel rapidly dephases in $T_{2,\uparrow}^*=4.1\pm0.7$~$\mu$s, while the coherence of the $\Ket{\downarrow}$ sublevel is strongly protected, persisting beyond the $50$~$\mu$s time frame of the measurement. This marks a $\gtrsim 19$-fold increase in $T_{2,\{m,p\}}^*$ over the bare $T_{2,\{+1,-1\}}^*$. We note that infidelities in our DQ pulses reduce the spin contrast within this subspace, limiting the utility of protecting only one sublevel in an unpolarized hyperfine manifold. Higher fidelity pulsing protocols or more efficient photon collection~\cite{siyushev2010} could increase the signal-to-noise ratio, which would make the lengthy coherence of the $\{\Ket{m,\downarrow},\Ket{p,\downarrow}\}$ qubit a valuable asset. 

In summary, we have experimentally demonstrated and theoretically analyzed the performance of mechanical CDD for decoupling an NV center spin qubit. We have shown that ac lattice strain can dress the spin states of an NV center and that the eigenstates of this dressed basis have robust coherence even in the presence of magnetic field fluctuations. We prolong $T_2^*$ of a thermally isolated qubit from $2.7\pm0.1$~$\mu$s to $15\pm1$~$\mu$s with a $\Omega/2\pi=581\pm2$~kHz mechanical dressing field and show that $T_2^*$ can be extended even further by either engineering more efficient devices or choosing to protect only a single $^{13}$C hyperfine sublevel. Mechanical CDD preserves the $\Ket{0}$ state and therefore does not require the NV center to be adiabatically dressed and undressed before and after each measurement. Moreover, the thermally sensitive $\{0,p\}$ and $\{0,m\}$ qubits maintain the gigahertz-scale Larmor frequency of their undressed analogs, providing rapid signal accumulation for a dressed state thermometer. Mechanically dressed qubits thus offer a promising option in the continuing development of NV center technology. 

\vspace*{4mm}

%
We thank J. Maxson, A. Bartnik, B. Dunham, and I. Bazarov at the Cornell University Cornell Laboratory for Accelerator-Based Sciences and Education (CLASSE) for electron irradiating the diamond sample used in this work for the creation of NV centers. We thank P. Maletinksy for interesting and useful discussions. Research support was provided by the Office of Naval Research (ONR) (Grant N000141410812). ERM received support from the Department of Energy Office of Science Graduate Fellowship Program (DOE SCGF), made possible in part by  the American Recovery and Reinvestment Act of 2009, administered by ORISE-ORAU under contract no. DE-AC05-06OR23100.  Device fabrication was performed in part at the Cornell NanoScale Science and Technology Facility, a member of the National Nanotechnology Coordinated Infrastructure, which is supported by the National Science Foundation (Grant ECCS-15420819), and at the Cornell Center for Materials Research Shared Facilities which are supported through the NSF MRSEC program (DMR-1120296). 

\section{Supplementary Information}

\section{Device Details}

We fabricated our device from an ``electronic grade'' $\langle100\rangle$-oriented diamond purchased from Element Six. The diamond is specified to contain fewer than $5$~ppb nitrogen impurities. Nitrogen-vacancy (NV) centers were introduced via irradiation with $2$~MeV electrons at a fluence of $\sim 1.2\times 10^{14}$~cm$^{-2}$ followed by annealing at $850^{\circ}$C for $2$~hours. The NV centers studied in this work are located at a depth of $\sim47$~$\mu$m. 

\begin{figure}[ht]
\begin{center}
\begin{tabular}{c}
\includegraphics[width=\linewidth]{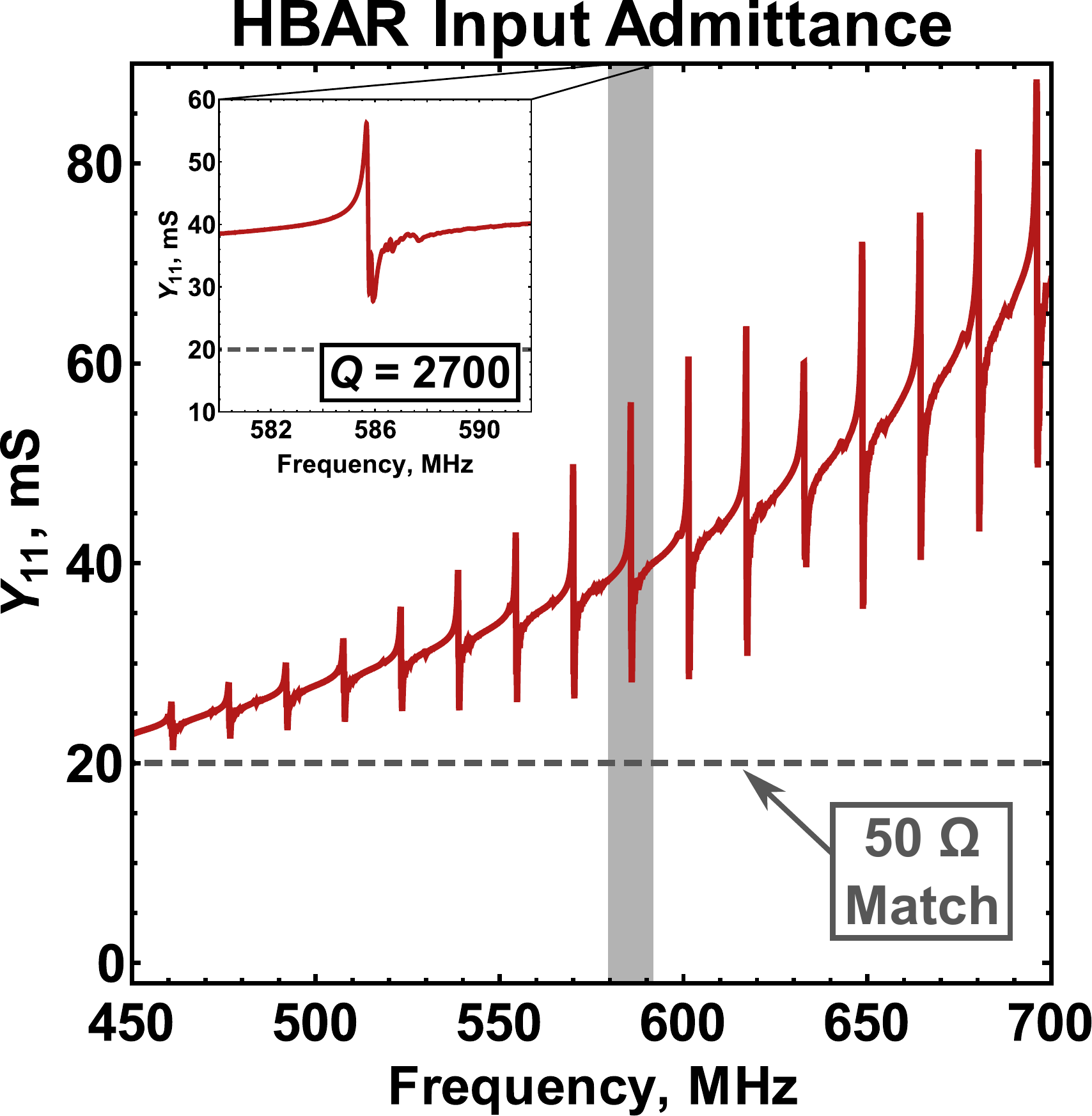} \\
\end{tabular} 
\end{center}
\caption[fig:s11] {Network analyzer measurement of the power admitted to the HBAR. The inset highlights the $\omega_{\text{mech}}/2\pi=586$~MHz mode used in the measurements.}
\label{fig:s11}
\end{figure}

The high-overtone bulk acoustic resonator (HBAR) used in these measurements consists of a $3$~$\mu$m thick $\langle002\rangle$-oriented ZnO film sandwiched between a Ti/Pt ($25$~nm/$200$~nm) ground plane and an Al ($250$~nm) top contact. The piezo-electric ZnO film transduces stress waves into the diamond. The diamond then acts as an acoustic Fabry-P\'{e}rot cavity to create stress standing wave resonances. Fig.~\ref{fig:s11} shows a network analyzer measurement of the HBAR admittance ($Y_{11}$) plotted as a function of frequency. From this frequency comb, we selected the $\omega_{\text{mech}}/2\pi=586$~MHz resonance mode that has a $Q$ of $2700$ as calculated by the $Q$-circle method~\cite{qCircle} and an on-resonance impedance of $18$~$\Omega$. This mechanical resonance suppresses driving field amplitude noise that is faster than $\omega_c=\frac{\omega_{\text{mech}}}{2Q}=110$~kHz. A microwave antenna fabricated on the diamond face opposite the ZnO transducer provides gigahertz frequency magnetic fields for conventional magnetic spin control. 

\section{Mechanical Rabi Driving}

\begin{figure}[ht]
\begin{center}
\begin{tabular}{c}
\includegraphics[width=\linewidth]{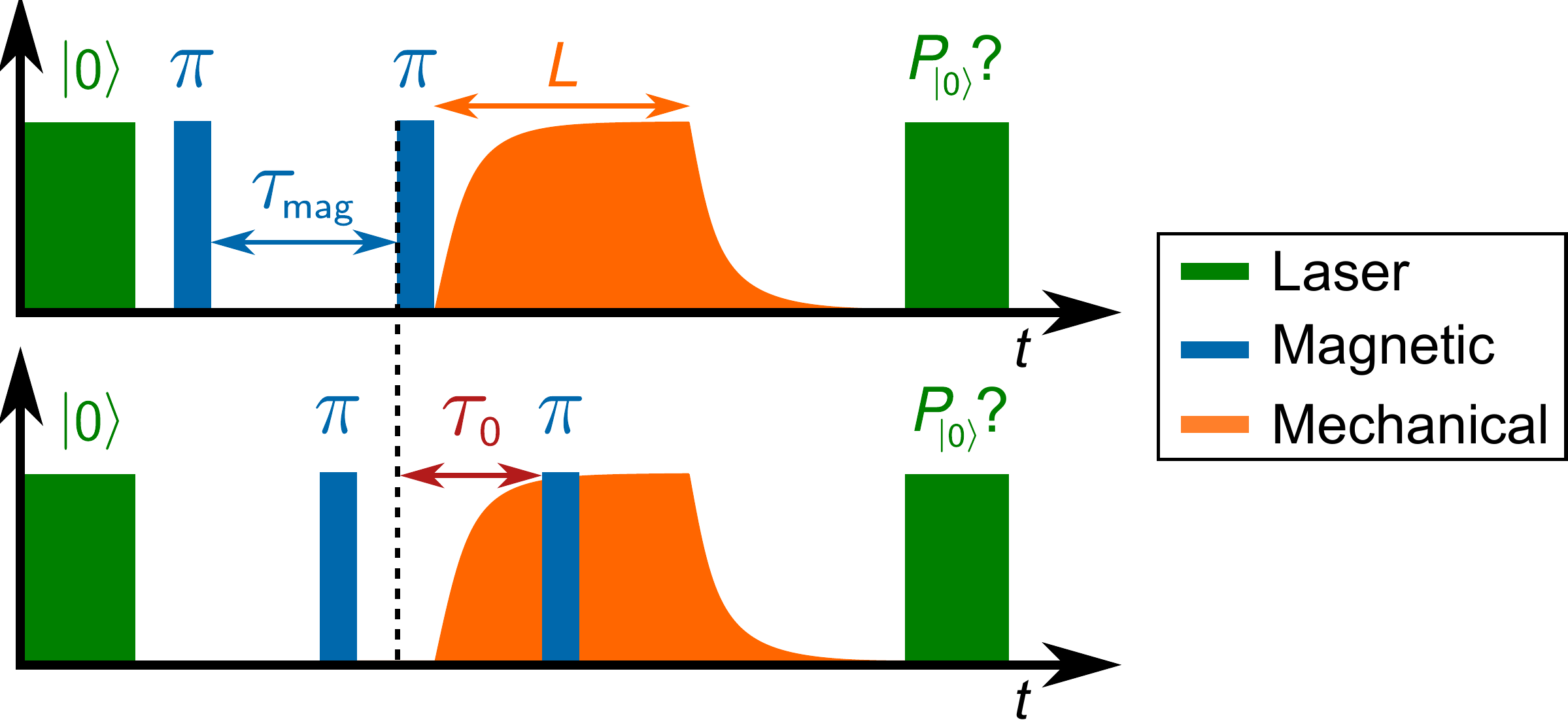} \\
\end{tabular} 
\end{center}
\caption[fig:rabiSeq]{Pulse sequence used to measure mechanically driven Rabi oscillations.}
\label{fig:rabiSeq}
\end{figure}

The mechanically driven Rabi oscillations depicted in Fig.~1a of the main text were measured using the pulse sequence shown in Fig.~\ref{fig:rabiSeq}. As described in detail in Ref.~\cite{MacQuarrie2015}, the relatively high $Q$ of our mechanical resonance makes it difficult to perform a traditional pulsed Rabi measurement. Instead, a pair of magnetic $\pi$-pulses resonant with the $\Ket{0}\leftrightarrow\Ket{-1}$ transition and separated by a fixed time $\tau_{\text{mag}}$ is swept through a fixed-length mechanical pulse. The mechanical pulse drives the $\Ket{+1}\leftrightarrow\Ket{-1}$ spin transition, and the duration of this interaction is set by the area of the mechanical pulse enclosed between the two $\pi$-pulses. By knowing the shape of the mechanical pulse, we convert this enclosed area to effective square-pulse units or an ``effective mechanical pulse length.'' Because the mechanical resonator is pulsed in this experiment, we are able to achieve a larger driving field than in the continuous dynamical decoupling (CDD) Ramsey measurements where the mechanical resonator operates in cw mode. 

\section{Mechanically Dressed Hamiltonian}
\label{sec:Ham}

As mentioned in the main text, we work within the $m_I=+1$ sublevel of the $^{14}$N hyperfine manifold. We consider both a static magnetic field $b$ that is aligned along the NV center symmetry axis and subject to fluctuations $\delta b$ and a mechanical driving field $\Omega$ that is subject to amplitude fluctuations $\delta\Omega$. In the $\{+1,0,-1\}\otimes\{(m_I=)+\frac{1}{2},-\frac{1}{2}\}$ Zeeman basis, a nearby $^{13}$C nuclear spin weakly couples to an NV center electronic spin through the hyperfine perturbation $H_{C}=A_{\|}S_{z}I_{z}$ where $S_z$ and $I_z$ are the spin-$1$ and spin-$\frac{1}{2}$ Pauli matrices, respectively, and $A_{\|}$ is the coupling strength~\cite{slichter}. An NV center electronic spin then obeys the Hamiltonian
\begin{widetext}
$H_{LF}=\begin{pmatrix}
\gamma b_{\Sigma}+\frac{1}{2}A_{\|} & 0 & 0 & 0 & \Omega_{\Sigma}\cos(\omega_{\text{mech}} t) & 0 \\
0 & \gamma b_{\Sigma}-\frac{1}{2}A_{\|} & 0 & 0 & 0 & \Omega_{\Sigma}\cos(\omega_{\text{mech}} t) \\
0 & 0 & -D & 0 & 0 & 0 \\
0 & 0 & 0 & -D & 0 & 0 \\
\Omega_{\Sigma}\cos(\omega_{\text{mech}} t) & 0 & 0 & 0 & -\gamma b_{\Sigma}-\frac{1}{2}A_{\|} & 0 \\
0 & \Omega_{\Sigma}\cos(\omega_{\text{mech}} t) & 0 & 0 & 0 & -\gamma b_{\Sigma}+\frac{1}{2}A_{\|} \\
\end{pmatrix}$
\end{widetext}
where $b_{\Sigma}=b+\delta b$, $\Omega_{\Sigma}=\Omega+\delta\Omega$, other parameters are as defined in the main text, and we have not included a magnetic driving field. Applying the rotating wave approximation and transforming into the reference frame rotating at $\frac{1}{2}\omega_{\text{mech}}=\frac{1}{2}(2\gamma b+\Delta)$ gives the Hamiltonian in the rotating frame
\begin{widetext}
$H_{RF}=
\begin{pmatrix}
\gamma b_{\Sigma}+\frac{1}{2}(\Delta+A_{\|}) & 0 & 0 & 0 & \frac{1}{2}\Omega_{\Sigma} & 0 \\
0 & \gamma b_{\Sigma}+\frac{1}{2}(\Delta+A_{\|}) & 0 & 0 & 0 & \frac{1}{2}\Omega_{\Sigma} \\
0 & 0 & -D & 0 & 0 & 0 \\
0 & 0 & 0 & -D & 0 & 0 \\
\frac{1}{2}\Omega_{\Sigma} & 0 & 0 & 0 & -\gamma b_{\Sigma}-\frac{1}{2}(\Delta+A_{\|}) & 0 \\
0 & \frac{1}{2}\Omega_{\Sigma} & 0 & 0 & 0 & -\gamma b_{\Sigma}-\frac{1}{2}(\Delta-A_{\|}) \\
\end{pmatrix}$.
\end{widetext}
Diagonalizing $H_{RF}$ gives the mechanically dressed Hamiltonian whose energies are quoted in the main text: 
\begin{widetext}
$H_{D}= \begin{pmatrix}
-D & 0 & 0 & 0 & 0 & 0 \\
0 & -D & 0 & 0 & 0 & 0 \\
0 & 0 & -\frac{1}{2}\sqrt{\Omega_{\Sigma}^2+\xi_{-}^2} & 0 & 0 & 0 \\
0 & 0 & 0 & \frac{1}{2}\sqrt{\Omega_{\Sigma}^2+\xi_{-}^2} & 0 & 0 \\
0 & 0 & 0 & 0 & -\frac{1}{2}\sqrt{\Omega_{\Sigma}^2+\xi_{+}^2} & 0 \\
0 & 0 & 0 & 0 & 0 & \frac{1}{2}\sqrt{\Omega_{\Sigma}^2+\xi_{+}^2} \\
\end{pmatrix}$
\end{widetext}
where $\xi_{\pm}=\Delta+2\gamma\delta b\pm A_{\|}$. In the limit $\Omega_{\Sigma}=0$, $H_D$ reduces to the undressed Zeeman Hamiltonian in the rotating frame. 

\section{Dressed State Spectroscopy}

\begin{figure}[ht]
\begin{center}
\begin{tabular}{c}
\includegraphics[width=\linewidth]{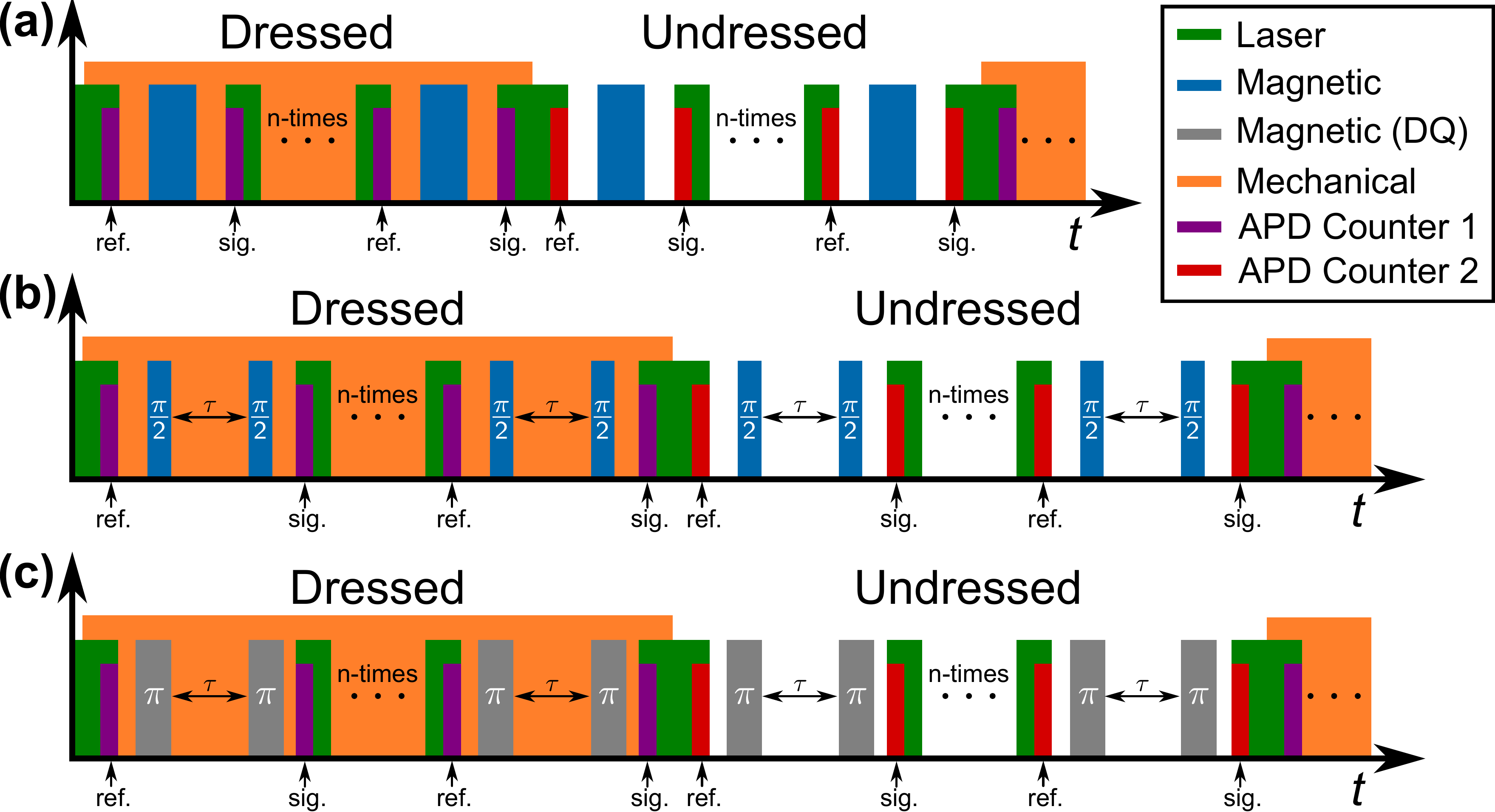} \\
\end{tabular} 
\end{center}
\caption[fig:seqs]{(a) Pulse sequence used for dressed state spectroscopy measurements. (b) Pulse sequence used for $\{0,p\}$ qubit CDD Ramsey measurements. (c) Pulse sequence used for $\{m,p\}$ qubit CDD Ramsey measurements. }
\label{fig:seqs}
\end{figure}

Fig.~\ref{fig:seqs}a shows several concatenated instances of the pulse sequence used for our dressed state spectroscopy measurements. In a single instance, the NV center is optically initialized into the $\Ket{0}$ spin state at which point a reference fluorescence measurement is made of the full-scale NV center photoluminescence. A magnetic $\pi$-pulse of strength $\Omega_{\text{mag}}/2\pi\sim80$~kHz is then applied to drive a conditional spin rotation. Finally, fluorescence readout provides a quantitative measure of the spin population remaining in $\Ket{0}$. We interleave $n$ instances of this pulse sequence executed in the dressed basis with $n$ instances of this pulse sequence executed in the undressed basis. In a typical experiment $n\sim10$, giving a total duty cycle time of $\sim280$~$\mu$s and mechanical pulse length of $\sim140$~$\mu$s. We differentiate between the dressed and undressed signal by routing the counts from our avalanche photodiode to separate counters on our DAQ. This sequence is then repeated as a function of the magnetic detuning $\Delta_{\text{mag}}$ from the $\Ket{0}\leftrightarrow\Ket{-1}$ state splitting to produce the data in Fig.~1d of the main text. 

The dressed signal from this measurement is fit to the sum of two Lorentzians 
\begin{equation}
\begin{split}
P_{D}&=c_D-\frac{a_{D,1}}{\left(\frac{2}{\Gamma_{D}}\right)^2\left(\omega-\frac{1}{2}\sqrt{\Delta^2+\Omega^2}-\frac{1}{2}\Delta-\omega_{0,-1}\right)^2+1} \\
& -\frac{a_{D,2}}{\left(\frac{2}{\Gamma_{D}}\right)^2\left(\omega+\frac{1}{2}\sqrt{\Delta^2+\Omega^2}-\frac{1}{2}\Delta-\omega_{0,-1}\right)^2+1}
\end{split}
\label{eq:specFit}
\end{equation}
where $P_{D}$ is the measured photoluminescence, $c_D$ is a constant background, $\omega_{0,-1}$ is the undressed $\Ket{0}\leftrightarrow\Ket{-1}$ spin state splitting, $\Delta$ is the mechanical detuning, $\Omega$ is the mechanical driving field, $a_{D,i}$ accounts for the depth of the spectral peaks, and $\Gamma_D$ is the full width at half maximum of the dressed spectral peaks. The undressed signal is simultaneously fit to the Lorentzian 
\begin{equation}
P_{UD}=c_{UD}-\frac{a_{UD}}{\left(\frac{2}{\Gamma_{UD}}\right)^2(\omega-\omega_{0,-1})^2+1}.
\end{equation}
We then subtract $\omega_{0,-1}$ from the $x$-axis to plot photoluminescence as a function of $\Delta_{mag}$ as shown in Fig.~1d of the main text.

\section{Expression for the Mechanical Detuning}

In our spectroscopy measurements, we use the relation $\frac{1}{2}(\omega_{0,m}+\omega_{0,p})-\omega_{0,-1}=\frac{1}{2}\Delta$ as a means of zeroing the mechanical detuning. To derive this expression, we begin in the $\{+1,0,-1\}$ basis with the Hamiltonian for an NV center subject to both a mechanical driving field and a magnetic driving field resonant with the $\Ket{0}\leftrightarrow\Ket{-1}$ transition. In the doubly rotating reference frame, this can be written
\begin{equation}
H_{RF}=\begin{pmatrix}
\frac{1}{2}\Delta & 0 & \frac{1}{2}\Omega \\
0 & -D-\Delta_{\text{mag}} & \frac{1}{2}\Omega_{\text{mag}} \\
\frac{1}{2}\Omega & \frac{1}{2}\Omega_{\text{mag}} & -\frac{1}{2}\Delta\\
\end{pmatrix}
\label{eq:hamDelta}
\end{equation}
where $\Delta_{\text{mag}}=-\frac{1}{2}\Delta$ for resonant magnetic driving, and $\Omega_{\text{mag}}$ is far enough detuned from the $\Ket{+1}\leftrightarrow\Ket{0}$ transition that we can ignore the $\Bra{+1}H_{RF}\Ket{0}$ matrix element. 

In the undressed case ($\Omega=0$, $\Delta=0$), the energy of the $\Ket{0}\leftrightarrow\Ket{-1}$ splitting in this reference frame is $\omega_{0,-1}=D$ where we define $\hbar=1$. With a non-zero mechanical driving field, calculating the eigenvalues of Eq.~\ref{eq:hamDelta} to first order in $\frac{\Omega_{\text{mag}}}{\Omega}$ gives energies $\omega_{0,p}=D+\frac{1}{2}(\Delta+\sqrt{\Delta^2+\Omega^2})$ and $\omega_{0,m}=D+\frac{1}{2}(\Delta-\sqrt{\Delta^2+\Omega^2})$. From this we arrive at the desired expression $\frac{1}{2}(\omega_{0,m}+\omega_{0,p})-\omega_{0,-1}=\frac{1}{2}\Delta$. The same expression is obtained when the $^{13}$C coupling is included. 

\section{Ramsey Measurements}

\subsection{Dressed Ramsey Pulse Sequences}

Fig.~\ref{fig:seqs}b,c show the pulse sequences used for our CDD Ramsey measurements of the $\{0,p\}$ and $\{m,p\}$ qubits, respectively. Similar to the spectroscopy experiments, the pulse sequences consist of $2n$ sub-instances where each sub-instance is a single measurement. Here, however, $n\sim 2$, which leads to mechanical pulse lengths and duty cycle lengths similar to those in the spectroscopy experiments. 

A single instance of the $\{0,p\}$ qubit CDD Ramsey sequence starts with optical initialization into $\Ket{0}$ and a reference fluorescence measurement. We then apply a magnetic $\pi/2$-pulse of strength $\Omega_{\text{mag}}/2\pi=696\pm 7$~kHz to populate the $\{0,p\}$ subspace. After a free evolution time $\tau$, we apply a second magnetic $\pi/2$-pulse of the same strength to return the spin population to $\Ket{0}$ where the signal is read out optically. To help visualize the decay, we advance the phase of the second $\pi/2$-pulse by $\omega_{\text{rot}}\tau$. Undressed Ramsey measurements are interleaved with the dressed measurements to reduce the power load on our device and provide a simultaneous measurement of the undressed dephasing time $T_{2,\{0,-1\}}^*$. This sequence is then repeated as a function of $\tau$.

The pulse sequence used for $\{m,p\}$ CDD Ramsey measurements is very similar to the $\{0,p\}$ Ramsey sequence. For the $\{m,p\}$ qubit, however, the $\pi/2$-pulses that address the $\{0,p\}$ subspace are replaced by double quantum magnetic $\pi$-pulses of strength $\Omega_{\text{mag}}/2\pi=1513\pm8$~kHz that address the $\{m,p\}$ subspace~\cite{mamin2014}. Additionally, the phase of the magnetic pulse that ends the free evolution time is not advanced at $\omega_{\text{rot}}\tau$ for the $\{m,p\}$ qubit measurement. In the interest of reducing the power load on our device, we interleave the dressed $\{m,p\}$ Ramsey measurements with undressed measurements that execute the same sequence of magnetic pulses. Because this pulse sequence amounts to a $2\pi$ rotation of the undressed $\{0,-1\}$ qubit, the data obtained during these measurements quantify the NV center spin contrast. For each measurement, the average of this undressed trace $\langle P_{0,ud}\rangle$ fixes the amplitude in the fitting functions described below. 

During the $\{m,p\}$ qubit measurements, we periodically measure $\Delta$ spectroscopically and feedback on $b$ to maintain a relatively constant $\Delta$. Interpolating linear drift between these measurements, we post-select to include only those data sets for which $\sigma_{\Delta}/2\pi<60$~kHz and $|\langle\Delta\rangle|/2\pi<35$~kHz. 

\subsection{Undressed Ramsey Fitting Function}

We fit the undressed Ramsey data to the expression 
\begin{equation}
\begin{split}
\text{Re}[\rho_{0,-1}]&=c-\frac{a}{4} e^{-\frac{\tau^2}{T_2^{*2}}}\{\cos[(\omega_{\text{rot}}+\Delta_{\text{mag}}+\frac{1}{2}A_{\|})\tau] \\
&+\cos[(\omega_{\text{rot}}+\Delta_{\text{mag}}-\frac{1}{2}A_{\|})\tau]\}
\end{split}
\end{equation}
where $\tau$ is the free evolution time, $\rho_{0,-1}$ is the $\{0,-1\}$ coherence, $c$ is a constant background, $a$ is an overall amplitude that accounts for deviations from perfect spin contrast, $T_2^*$ is the inhomogeneous dephasing time, $\omega_{\text{rot}}$ is the rate at which we advance the phase of the second $\pi/2$-pulse, $\Delta_{\text{mag}}$ is the magnetic detuning, and $A_{\|}$ quantifies coupling to a nearby $^{13}$C nuclear spin. Of these values, $c$, $a$, $T_{2}^*$, $\Delta_{\text{mag}}$, and $A_{\|}$ are free parameters in our fit. We have assumed the $^{13}$C spin is unpolarized. We use the values of $a$ and $c$ returned from the fits to scale the $y$-axes of our plots. 

\subsection{Dressed Ramsey Fitting Function: The $\{0,p\}$ Qubit}

In our CDD Ramsey measurements of the $\{0,p\}$ qubit, we tune the magnetic driving field into resonance with the $\Ket{0}\leftrightarrow\Ket{p}$ transition. For the fits, we zero the magnetic detuning midway between the $^{13}$C sublevels $\Ket{p,(m_{I}=)+\frac{1}{2}}$ and $\Ket{p,-\frac{1}{2}}$. Assuming $\Delta=0$, our $\{0,p\}$ CDD Ramsey signal is then described by the expression
\begin{equation}
\begin{split}
\text{Re}[\rho_{0,p}]&=c+\frac{1}{4}e^{-\frac{\tau^2}{T_2^{*2}}}\lbrace a_p \cos\left[\left(\Delta_{\text{mag}}+\omega_{\text{rot}}\right)\tau+\phi\right] \\
&+a_m \cos\left[\left(\Delta_{\text{mag}}+\omega_{\text{rot}}+\sqrt{\Omega^2+A_{\|}^2}\right)\tau+\phi\right]\rbrace
\end{split}
\end{equation}
where $a_m$ is the spin contrast for the $\{0,m\}$ qubit, $a_p$ is the spin contrast for the $\{0,p\}$ qubit, $\phi$ is a constant phase offset, and the other parameters are as defined above. We fix the values of $A_{\|}$ and $\omega_{\text{rot}}$, and we vary $c$, $a_i$, $\phi$, $\Omega$, and $\Delta_{\text{mag}}$ as free parameters in our fitting procedure. Once again, we use the values of $a_m$, $a_p$, and $c$ returned from the fit to scale the $y$-axis in Fig.~2a of the main text. 

It is important to note that because the dressed qubit Larmor frequency does not scale linearly with magnetic field fluctuations, the Ramsey signal does not follow a strictly Gaussian decay. Nevertheless, we fit our data with a Gaussian envelope to aid comparison with the undressed dephasing time. This is a reasonable approximation over the range of mechanical driving fields accessed in this work. 

\subsection{Dressed Ramsey Fitting Function: The $\{m,p\}$ Qubit}

\begin{figure}[ht]
\begin{center}
\begin{tabular}{c}
\includegraphics[width=\linewidth]{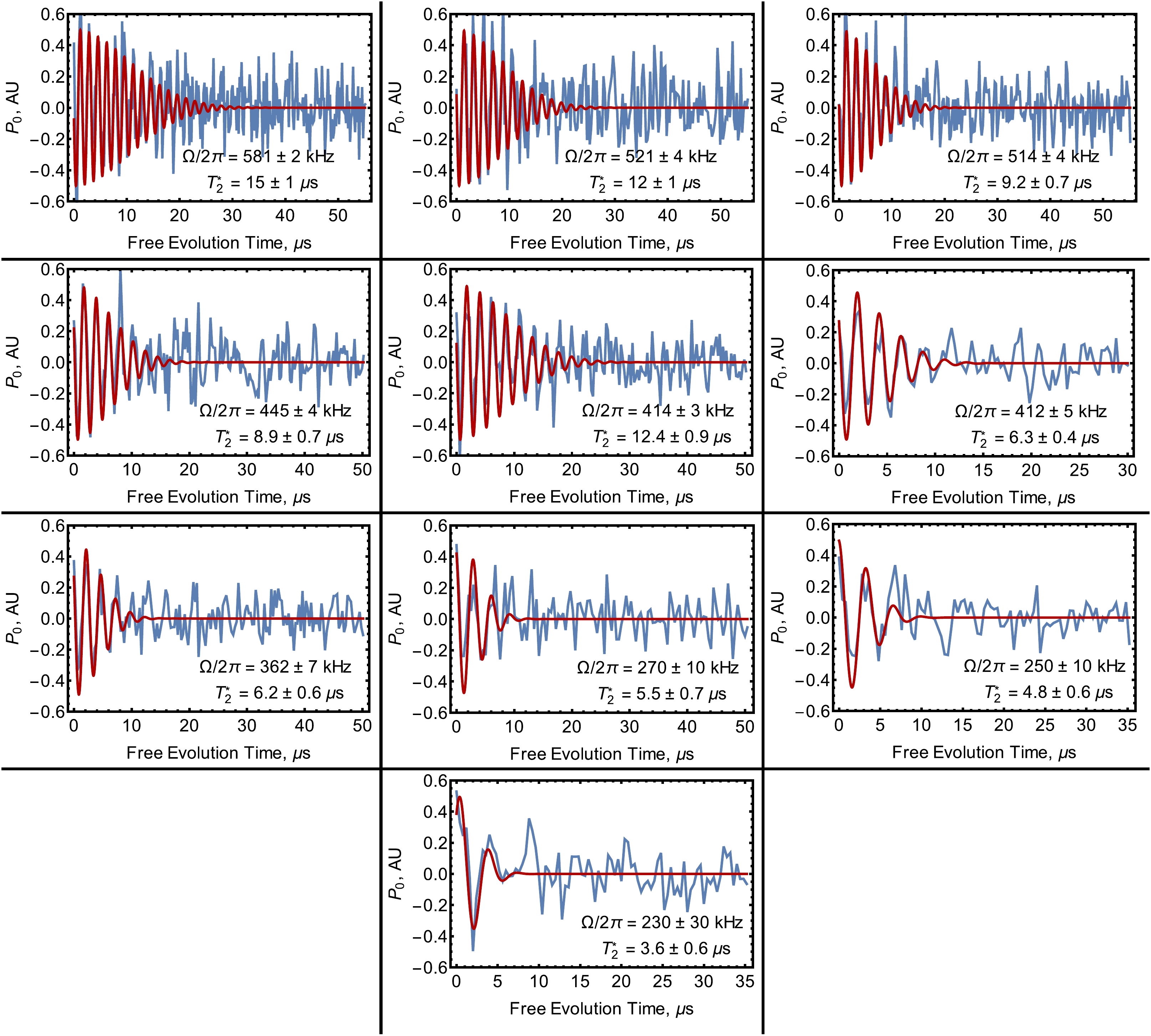} \\
\end{tabular} 
\end{center}
\caption[fig:tablePlot]{Data and fits for CDD Ramsey measurements of the $\{m,p\}$ qubit when $\Omega$ was actively stabilized. }
\label{fig:tablePlot}
\end{figure}

\begin{figure}[ht]
\begin{center}
\begin{tabular}{c}
\includegraphics[width=\linewidth]{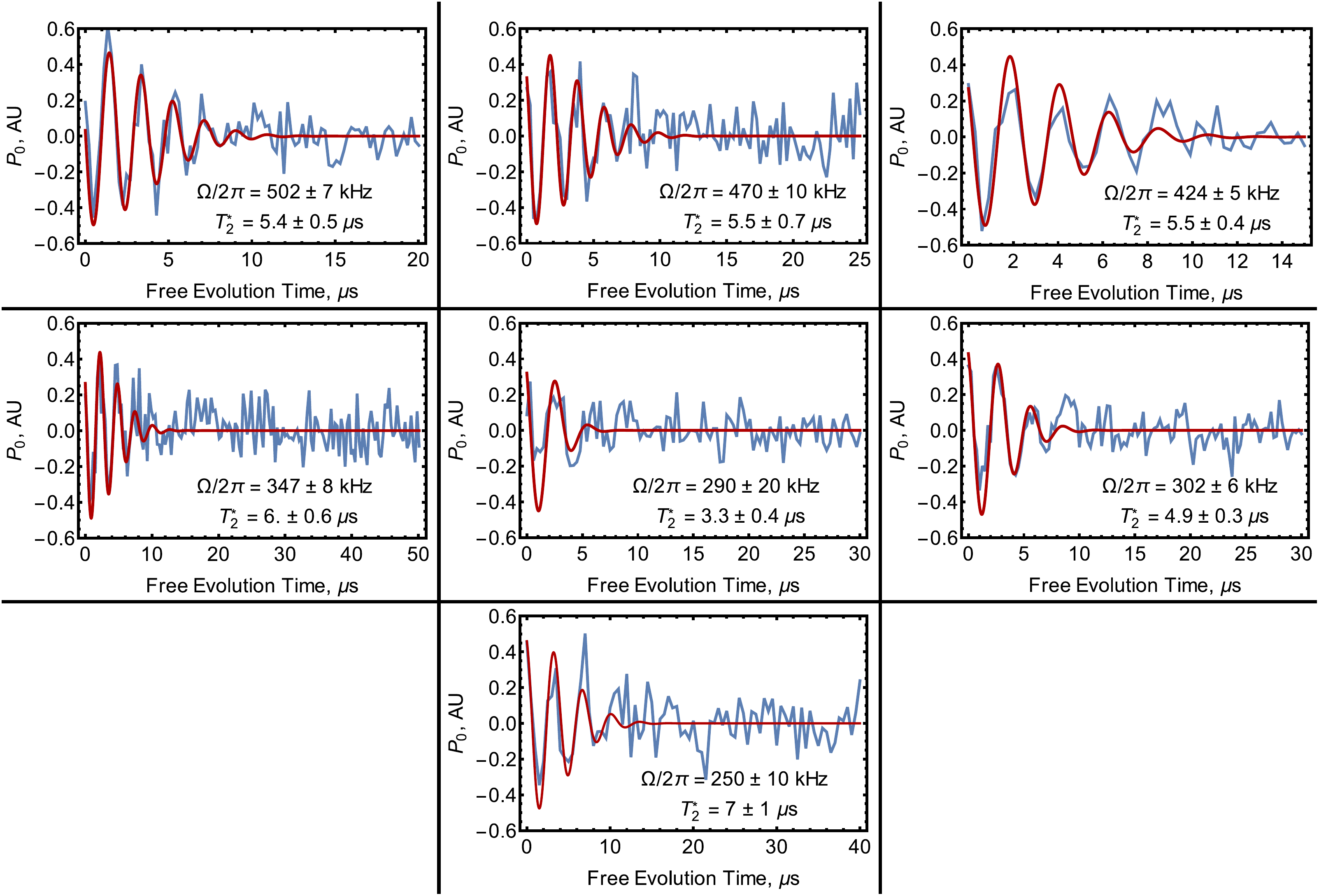} \\
\end{tabular} 
\end{center}
\caption[fig:tablePlotN]{Data and fits for CDD Ramsey measurements of the $\{m,p\}$ qubit when $\Omega$ was given a Gaussian noise profile. }
\label{fig:tablePlotN}
\end{figure}

For the $\{m,p\}$ qubit dressed under the condition $\Delta=0$, our CDD Ramsey signal can be described by the expression
\begin{equation}
\text{Re}[\rho_{m,p}]=c+\frac{\langle P_{0,ud}\rangle}{2}e^{-\frac{\tau^2}{T_2^{*2}}} \cos\left[ \tau\sqrt{A_{\|}^2+\Omega^2}+\phi\right]
\end{equation}
where $\langle P_{0,ud}\rangle$ measures the spin contrast and the other parameters are as described above. To maximally constrain our fitting procedure, we measure $\langle P_{0,ud}\rangle$ by interleaving undressed iterations of the CDD Ramsey protocol into the measurement. We allow $c$, $T_{2}^*$, $\Omega$, and $\phi$ to vary as free parameters in our fitting procedure. The results of these fits for the measurements shown in Fig.~3b,c of the main text are displayed in Fig.~\ref{fig:tablePlot} and Fig.~\ref{fig:tablePlotN}, respectively. 

Our CDD Ramsey measurement of a maximally protected $\Ket{\downarrow}$ $^{13}$C sublevel (Fig.~3d in the main text) was fit to the function 
\begin{equation}
\begin{split}
\text{Re}[\rho_{m,p}]&=\frac{\langle P_{0,ud}\rangle}{4}\lbrace\sqrt{\frac{\Omega}{\sqrt{\Omega^2+(2\gamma\sigma_b)^4\tau^2}}}\cos\left[\Omega\tau+\phi\right] \\
&+ e^{-\frac{t^2}{T_{2,\uparrow}^*2}}\cos\left[\tau\sqrt{\Omega^2+4A_{\|}^2}+\phi\right]\rbrace +c
\end{split}
\label{eq:highCohere}
\end{equation}
where $\langle P_{0,ud}\rangle$ fixes the spin contrast, and $\Omega$, $\phi$, $T_{2,\uparrow}^*$, and $c$ were varied as free parameters. A derivation of the non-Gaussian envelope in Eq.~\ref{eq:highCohere} is given below on page~\pageref{sndOrder}. 

For all of our $\{m,p\}$ qubit Ramsey plots, we use $\langle P_{0,ud} \rangle$ and the value of $c$ returned from the fit to scale the $y$-axis. 

\section{Thermal Stability}

As mentioned above, we intersperse spectral measurements within CDD Ramsey measurements of the $\{m,p\}$ qubit. This allows us to feedback on $b$ and maintain a relatively constant $\Delta$, but these measurements also quantify the thermal drift over the course of the measurement. A histogram of $\Delta$ extracted from fitting these spectra to Eq.~\ref{eq:specFit} quantifies drift in the magnetic bias field $\sigma_{\Delta}=2\gamma\sigma_{\text{bias}}$. A histogram of $\omega_{0,-1}$, however, provides information about both the magnetic bias field drift and the thermal drift according to 
\begin{equation}
\sigma_{0,-1}=\sqrt{\left(\gamma\sigma_{\text{bias}}\right)^2+\left(\frac{dD}{dT}\sigma_T\right)^2}
\end{equation}
where $\sigma_T$ is the standard deviation of normally distributed thermal drift and $\frac{dD}{dT}=-74\times 2\pi$~kHz/$^{\circ}$C is the temperature dependence of $D$~\cite{awschalomThermo,budkerThermo}. The average of $\sigma_T$ for the power-leveled data that satisfy our post-selection criteria is $0.25\pm 0.03^{\circ}$C. Thermal drift on a similar scale can be expected for the $\{0,p\}$ qubit measurements. As shown below on page~\pageref{tempCoh}, fluctuations of this scale would limit the $\{0,p\}$ qubit coherence time to $T_{2,\{0,p\}}^*=\frac{\sqrt{2}}{\sigma_{T} dD/dT}=12\pm 1$~$\mu$s. 

\section{Modeling Decoherence}

\subsection{First Order Fluctuations in $\omega_{i,j}$}

Generically, first order deviations in the Larmor frequency $\omega_{i,j}$ take the form $\delta\omega_{i,j}=\alpha\delta x$ where $\alpha$ is a constant. If the fluctuation $\delta x$ follows a Gaussian distribution with standard deviation $\sigma_x$, an expression for the associated dephasing rate can be found by calculating the weighted average of a distribution of detuned, un-damped Ramsey signals:
\begin{equation}
\begin{split}
\text{Re}[\rho_{i,j}] & =\frac{1}{\sqrt{2\pi}\sigma_x}\int e^{-\frac{\delta x^2}{2\sigma_x^2}}\cos\left[(\omega_{i,j}+\alpha\delta x)\tau\right]\text{d}\delta x \\
& = e^{-\frac{1}{2}(\alpha\sigma_x \tau)^2}\cos\left(\omega_{i,j}\tau\right).
\end{split}
\label{eq:sigReplace}
\end{equation}
Comparing Eq.~\ref{eq:sigReplace} with an ideal Ramsey signal given by $\text{Re}[\rho_{i,j}]=e^{-\frac{\tau^2}{T_{2,\{i,j\}}^{*2}}}\cos\left(\omega_{i,j}\tau\right)$, we see that $T_{2,\{i,j\}}^*=\frac{\sqrt{2}}{\alpha \sigma_x}$ and therefore $\Gamma_x=\frac{2\pi}{T_{2,\{i,j\}}^*}=\sqrt{2}\pi\alpha \sigma_x$. 

For magnetic field fluctuations experienced by the $\{0,-1\}$ qubit, $\alpha\delta x\rightarrow \gamma\delta b$. We then find $\gamma\sigma_b/2\pi=(\sqrt{2}\pi T_{2,\{0,-1\}}^*)^{-1}$ as quoted in the main text. For thermal fluctuations experienced by the $\{0,p\}$ qubit, $\alpha\delta x\rightarrow \frac{\text{d}D}{\text{d}T}\delta T$, and we arrive at $T_{2,\{0,p\}}^*=\frac{\sqrt{2}}{\sigma_{T} dD/dT}$. For the $\{m,p\}$ qubit, expanding $\omega_{m,p}$ to first order in $\delta b$ gives 
\begin{equation}
\delta\omega_{m,p;b}=\frac{2|A_{\|}|\gamma\delta b}{\sqrt{A_{\|}^2+\Omega^2}},
\end{equation}
from which we find $\Gamma_b=2\sqrt{2}\kappa|A_{\|}|/T_{2,\{0,-1\}}^*$ where $\frac{1}{\kappa}=\frac{1}{\sqrt{2}\pi}\sqrt{A_{\|}^2+\Omega^2}$. Similarly, expanding $\omega_{m,p}$ to first order in $\delta \Omega$ gives
\begin{equation}
\delta\omega_{m,p;\Omega}=\frac{\Omega \delta\Omega}{\sqrt{A_{\|}^2+\Omega^2}},
\end{equation}
from which we find $\Gamma_\Omega=\kappa\Omega\sigma_{\Omega}$. 
\label{tempCoh}

\subsection{Second Order Magnetic Field Fluctuations}
\label{sndOrder}

The decay envelope of a Ramsey measurement is given by the expression $f(\tau,\Omega,\sigma_b,A_{\|})=|\langle e^{i\delta\phi}\rangle|$ where $\delta\phi$ is the random phase accumulated in a given duty cycle of the measurement~\cite{ithier2005}. For the $\{m,p\}$ qubit in the case when $\Delta=0$, the Larmor frequency is given by $\omega_{m,p}=\sqrt{(\Omega+\delta\Omega)^2+(A_{\|}+2\gamma b)^2}$. To second order in $\delta b$, fluctuations in $\omega_{m,p}$ from magnetic field fluctuations are then given by 
\begin{equation}
\begin{split}
\delta\omega_{m,p}&=\frac{\partial\omega_{m,p}}{\partial b}\Bigr|_{\delta b=0}\delta b+\frac{\partial^2\omega_{m,p}}{\partial b^2}\Bigr|_{\delta b=0}\frac{\delta b^2}{2}+O(\delta b^3) \\
&=\frac{2\gamma\delta b(A_{\|}^3+A_{\|}\Omega^2+\gamma\delta b\Omega^2)}{(A_{\|}^2+\Omega^2)^{3/2}}
\end{split}.
\end{equation}
The random phase accumulated is $\delta\phi=\delta\omega_{m,p}\tau$. By averaging this phase over a Gaussian distribution of magnetic field fluctuations, we find 
\begin{equation}
\begin{split}
f(\tau,\Omega,\sigma_b,A_{\|})&=\left\vert\frac{1}{\sqrt{2\pi}\sigma_b}\int_{-\infty}^{\infty}e^{i\delta\omega_{m,p}\tau}e^{-\frac{\delta b^2}{2\sigma_b^2}}\text{d}\delta b\right\vert \\
&=\sqrt{\beta}e^{-\frac{2(\gamma\sigma_b A_{\|}\beta\tau)^2}{A_{\|}^2+\Omega^2}}
\end{split}
\label{eq:env}
\end{equation}
where
\begin{equation}
\beta(\tau,\Omega,\sigma_b,A_{\|})\equiv\sqrt{\frac{(A_{\|}^2+\Omega^2)^3}{(A_{\|}^2+\Omega^2)^3+(2\gamma\sigma_b\Omega)^4\tau^2}}.
\end{equation}
To produce the model curves in Fig.~3b,c of the main text, we numerically solve this expression for the value of $\tau$ such that $f(\tau,\Omega,\sigma_b,A_{\|})=\frac{1}{e}$. 

When $\Delta=-|A_{\|}|$, the two $^{13}$C sublevels follow different decay envelopes that can be computed by setting $A_{\|}\rightarrow 0$ and $A_{\|}\rightarrow 2 A_{\|}$ in Eq.~\ref{eq:env}. In the former case, $f(\tau,\Omega,\sigma_b,0)$ reduces to 
\begin{equation}
\begin{split}
h(\tau,\Omega,\sigma_b)&=\sqrt{\frac{\Omega}{\sqrt{\Omega^2+(2\gamma\sigma_b)^4\tau^2}}}
\end{split}
\end{equation}
as seen in the main text. For the case of $A_{\|}\rightarrow 2 A_{\|}$, we approximate the decay as Gaussian. The fitting function for Fig.~3d of the main text then becomes
\begin{equation}
\begin{split}
\text{Re}[\rho_{m,p}]&=c+\frac{\langle P_{0,ud}\rangle}{4}\lbrace\sqrt{\frac{\Omega}{\sqrt{\Omega^2+(2\gamma\sigma_b)^4\tau^2}}}\cos\left[\Omega\tau+\phi\right] \\
&+e^{-\frac{t^2}{T_{2,\uparrow}^{*2}}}\cos\left[\tau\sqrt{\Omega^2+4 A_{\|}^2}+\phi\right]\rbrace
\end{split}
\end{equation}
where only $\Omega$, $\phi$, $c$, and $T_{2,\uparrow}^*$ were allowed to vary as free parameters.

For simplicity, this derivation of $f(\tau,\Omega,\sigma_b,A_{\|})$ does not include driving field noise. Including amplitude noise in the mechanical driving field on the scale of our power-leveled measurements produces no noticeable change in the results of the model over the range of mechanical driving fields addressed here. 

\section{Measuring the Voltage Reflected from the HBAR}

\begin{figure}[ht]
\begin{center}
\begin{tabular}{c}
\includegraphics[width=\linewidth]{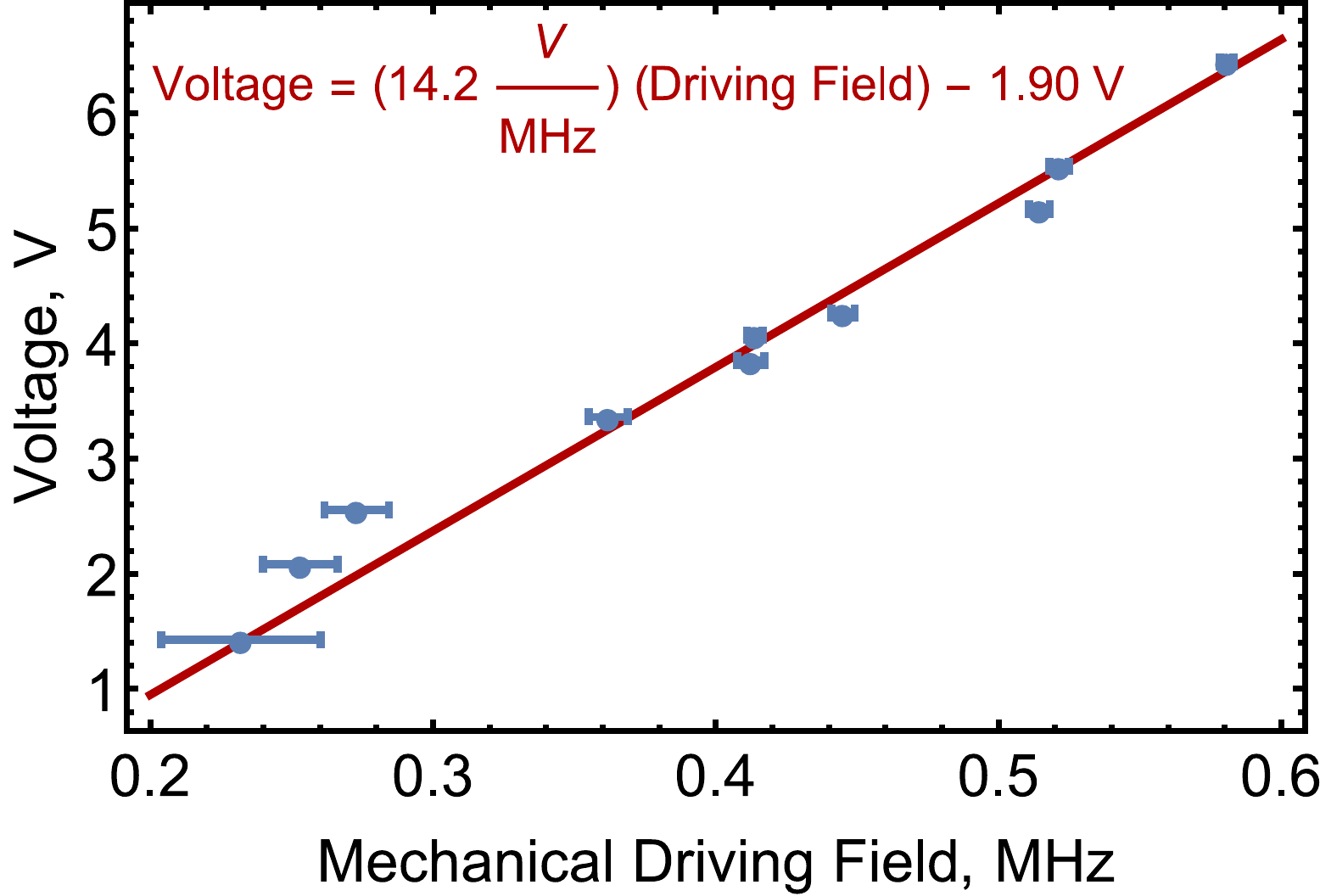} \\
\end{tabular} 
\end{center}
\caption[fig:diodecal] {Voltage reflected from the mechanical resonator plotted as a function of the mechanical driving field. }
\label{fig:diodecal}
\end{figure}

We monitor the mechanical driving field amplitude by tracking the RF power reflected from the mechanical resonator. An RF circulator redirects the reflected power to an RF diode that converts the ac signal to the dc voltage that we measure. As shown in Fig.~\ref{fig:diodecal}, this measured voltage scales linearly with the mechanical driving field. However, due to the diode's nonzero threshold voltage, that linear dependence has a nonzero intercept. 

We introduce driving field noise to our experiment by periodically shifting the applied power such that the spread of voltages measured by the RF diode over the course of a measurement is normally distributed with a standard deviation of $\eta \langle V_R \rangle$ where $V_R$ is the reflected voltage and $\eta$ is a constant. Because Fig.~\ref{fig:diodecal} has a nonzero intercept, such a distribution of voltages will correspond to a Gaussian distribution of driving fields with a standard deviation of $\sigma_{\Omega}=(\langle\Omega\rangle+\alpha)\eta$ where $\alpha/2\pi=-133\pm7$~kHz is the ratio of the intercept to the slope for the line of best fit in Fig.~\ref{fig:diodecal}. 

\section{Coherence of the $\{+1,-1\}$ Qubit}

\begin{figure}[ht]
\begin{center}
\begin{tabular}{c}
\includegraphics[width=\linewidth]{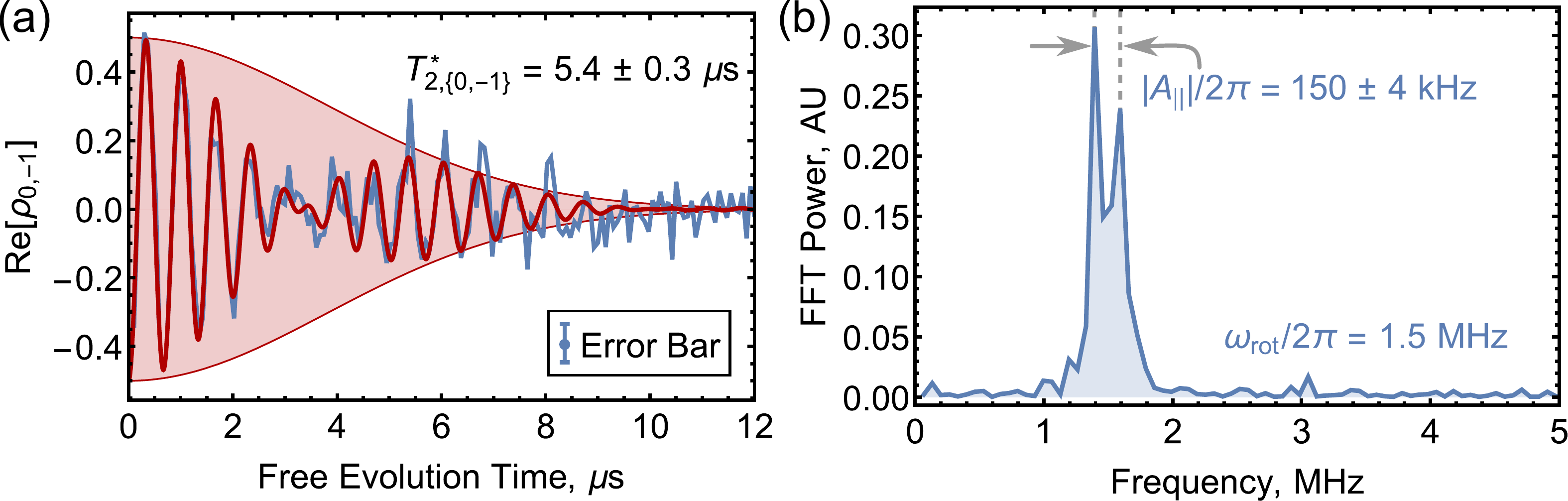} \\
\end{tabular} 
\end{center}
\caption[fig:t2UD]{(a) Ramsey measurement of the undressed $\{0,-1\}$ qubit for the NV center used in the $\{m,p\}$ qubit measurements. (b) Fourier spectrum of (a). }
\label{fig:t2UD}
\end{figure}

We compare the coherence of the $\{m,p\}$ qubit to that of the undressed $\{+1,-1\}$ qubit because in each of these qubits both component states are sensitive to magnetic field fluctuations. Directly measuring the dephasing time of the $\{+1,-1\}$ qubit at finite field with high precision is a non-trivial task because the measurement becomes sensitive to double quantum pulse infidelities. Instead, we measure $T_2^*$ of the undressed $\{0,-1\}$ qubit (Fig.~\ref{fig:t2UD}) and rely on the fact that for Gaussian magnetic field fluctuations $T_{2,\{+1,-1\}}^*=\frac{1}{2}T_{2,\{0,-1\}}^*$. This gives $T_{2,\{+1,-1\}}^*=2.7\pm0.1$~$\mu$s as quoted in the main text. This same undressed Ramsey measurement also quantifies $|A_{\|}|/2\pi=150\pm4$~kHz and $\sigma_b=2.4\pm0.1$~mG for this NV center.

\section{Dressed Spectra Through the $\Ket{+1}\leftrightarrow\Ket{0}$ Transition}

\begin{figure}[hb]
\begin{center}
\begin{tabular}{c}
\includegraphics[width=\linewidth]{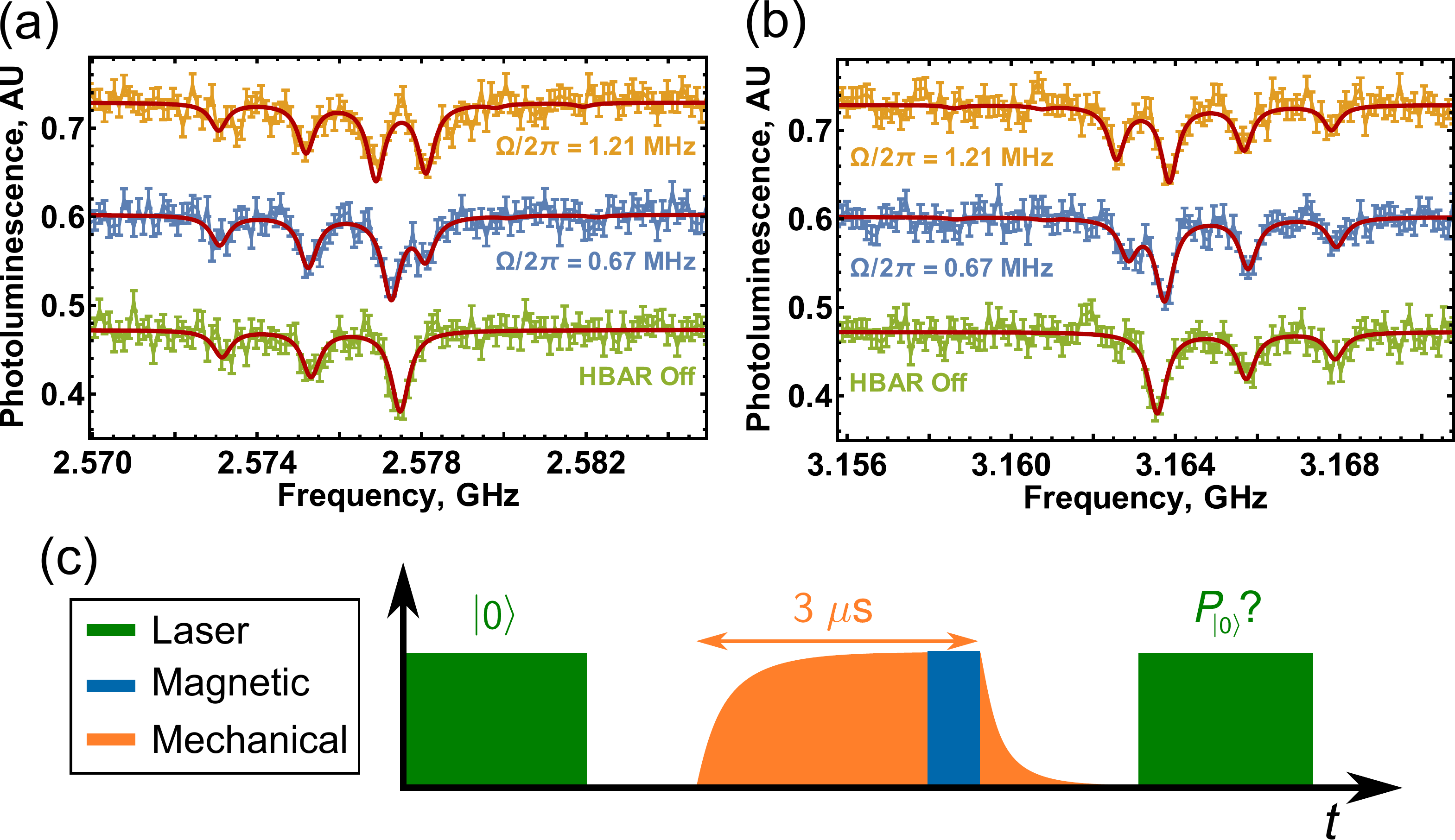} \\
\end{tabular} 
\end{center}
\caption[fig:0pSpect]{(a) Dressed state spectrum for a magnetic pulse swept through the undressed $\Ket{0}\leftrightarrow\Ket{-1}$ transition. (b) Dressed state spectrum for a magnetic pulse swept through the undressed $\Ket{+1}\leftrightarrow\Ket{0}$ transition. (c) Pulse sequence used in these measurements.}
\label{fig:0pSpect}
\end{figure}

Fig.~\ref{fig:0pSpect} shows spectral measurements of the dressed state splitting as measured by sweeping the detuning of a $\Omega_{\text{mag}}/2\pi=345\pm4$~kHz magnetic pulse through the resonance of the undressed (a) $\Ket{0}\leftrightarrow\Ket{-1}$ and (b) $\Ket{+1}\leftrightarrow\Ket{0}$ transitions. All three $^{14}$N hyperfine sublevels are visible in the spectra. Because $\omega_{\text{mech}}$ is tuned into resonance with the $\Ket{(m_{s}=)+1,(m_{I}=)+1}\leftrightarrow\Ket{-1,+1}$ transition within the $^{14}$N hyperfine manifold, only the $m_{I}=+1$ peak splits into the dressed states $\Ket{m,+1}$ and $\Ket{p,+1}$. In these measurements, the HBAR was powered in $3$~$\mu$s pulses as shown in Fig.~\ref{fig:0pSpect}c. This reduced the power load on the device and allowed us to reach higher driving fields than we were able to reach in the CDD Ramsey experiments where the mechanical resonator operates in cw mode. 

\bibliography{bibDiMEMS}

\end{document}